\documentclass[11pt]{article}
\usepackage{amssymb,amsmath,amsfonts}
\usepackage{graphicx}
\usepackage{graphics}
\usepackage{eepic,epsfig}

\begin{document}

\title{Casimir effect for parallel plates in de Sitter spacetime}
\author{E. Elizalde$^{1}$, A.~A. Saharian$^{2}$, T.~A. Vardanyan$^{2}$ \\
\\
\textit{$^{1}$Instituto de Ciencias del Espacio (CSIC) }\\
\textit{and Institut d'Estudis Espacials de Catalunya (IEEC/CSIC) }\\
\textit{Campus UAB, Facultat de Ci\`{e}ncies, Torre C5-Parell-2a planta,}\\
\textit{08193 Bellaterra (Barcelona) Spain}\vspace{0.3cm}\\
$^{2}$\textit{Department of Physics, Yerevan State University,}\\
\textit{1 Alex Manoogian Street, 0025 Yerevan, Armenia}}
\maketitle

\begin{abstract}
The Wightman function and the vacuum expectation values of the field squared
and of the energy-momentum tensor are obtained, for a massive scalar field
with an arbitrary curvature coupling parameter, in the region between two
infinite parallel plates, on the background of de Sitter spacetime. The
field is prepared in the Bunch--Davies vacuum state and is constrained to
satisfy Robin boundary conditions on the plates. For the calculation, a
mode-summation method is used, supplemented with a variant of the
generalized Abel-Plana formula. This allows to explicitly extract the
contributions to the expectation values which come from each single
boundary, and to expand the second-plate-induced part in terms of
exponentially convergent integrals. Several limiting cases of interest are
then studied. Moreover, the Casimir forces acting on the plates are
evaluated, and it is shown that the curvature of the background spacetime
decisively influences the behavior of these forces at separations larger
than the curvature scale of de Sitter spacetime. In terms of the curvature
coupling parameter and the mass of the field, two very different regimes are
realized, which exhibit monotonic and oscillatory behavior of the vacuum
expectation values, respectively. The decay of the Casimir force at large
plate separation is shown to be power-law (monotonic or oscillating), with
independence of the value of the field mass.
\end{abstract}

\bigskip


\bigskip

\section{Introduction}

The Casimir effect \cite{Casimir} is now known to be common to systems of
very different kind, involving fluctuating quantities on which external
boundary conditions are imposed. It can have important implications on all
scales, from subnuclear to cosmological. Imposing boundary conditions on a
quantum field leads to a modification of the spectrum of zero-point
fluctuations and results in the shifting in the vacuum expectation values
for physical quantities, such as the energy density and stresses. In
particular, the confinement of quantum fluctuations induces forces that act
on the constraining boundaries. The particular features of the resulting
vacuum forces depend on the nature of the quantum field, on the type of the
spacetime manifold, the geometry of the boundaries, and on the specific
boundary conditions imposed on the field.

An interesting topic in the investigation of the Casimir effect is its
explicit dependence on the geometry of the background spacetime. As usual,
the relevant information is encoded in the vacuum fluctuations spectrum and,
not surprisingly, analytic solutions can be found for highly symmetric
geometries only. In special, motivated by Randall--Sundrum type braneworld
scenarios, investigations of the Casimir effect in anti-de Sitter (AdS)
spacetime have attracted a great deal of attention. The braneworld
corresponds to a manifold with boundaries and all fields which propagate in
the bulk will give Casimir-type contributions to the vacuum energy and, as a
result, to the vacuum forces acting on the branes. The Casimir effect
provides in this context a natural mechanism for stabilizing the radion
field in the Randall--Sundrum model, as required for a complete solution of
the hierarchy problem. In addition, the Casimir energy gives a contribution
to both the brane and bulk cosmological constants and, hence, it has to be
taken into account in any self-consistent formulation of the braneworld
dynamics. The Casimir energy and corresponding Casimir forces for two
parallel branes in AdS spacetime have been evaluated in Refs.~\cite{Gold00}
by using either dimensional or zeta function regularization methods. Local
Casimir densities were considered in Refs.~\cite{Knap04,Saha05}. The Casimir
effect in higher-dimensional generalizations of the AdS spacetime with
compact internal spaces has been investigated in~\cite{Flac03}.

Another popular background in gravitational physics is de Sitter (dS)
spacetime. Quantum field theory in this background has been extensively
studied during the past two decades. Much of the early interest was
motivated by questions related with the quantization of fields propagating
on curved backgrounds. The dS spacetime has a high degree of symmetry and
numerous physical problems are exactly solvable on this background.
Importance of such theoretical work was increased with the appearance of the
inflationary cosmology scenario~\cite{Lind90}. In most inflationary models,
an approximate dS spacetime is employed with the aim to solve a number of
problems in standard cosmology. During the inflationary epoch, quantum
fluctuations in the inflaton field introduce inhomogeneities which play a
central role in the generation of cosmic structures from inflation. More
recently, astronomical observations of standard-candle supernovae, galaxy
clusters and the cosmic microwave background \cite{Ries07} have clearly
indicated that at present our (local) universe is accelerating and can be
well approximated by $\Lambda$CDM, FRW cosmology with a positive
cosmological constant $\Lambda$. Now, if the universe, as it seems, is going
to accelerate for ever, this cosmology will lead asymptotically to a dS
universe. Another motivation for the investigation of dS-based quantum
theories is related to the holographic duality known to hold between quantum
gravity on dS spacetime and a quantum field theory living on its boundary,
identified with the timelike infinity surface of the dS spacetime.

Motivated by the above considerations---and since they are ingredients that
more full-fledged models will necessarily have to incorporate---we will here
calculate the Casimir densities and forces arising for the geometry of two
parallel plates on the background of $(D+1)$-dimensional dS spacetime.
Previously, the Casimir effect on the background of dS spacetime described
in planar coordinates was investigated in Refs.~\cite{Seta01} for a
conformally coupled massless scalar field. In this last case the problem is
conformally related to the corresponding problem in Minkowski spacetime and
the vacuum characteristics are generated from those for the Minkowski
counterpart, just by multiplying with the conformal factor. In particular,
for the geometry of a single plate, the vacuum expectation value of the
energy-momentum tensor vanishes. The Casimir density induced by a single
plate for a massive scalar field with an arbitrary curvature coupling
parameter has been considered in \cite{Saha09}. In \cite{Saha04} the vacuum
expectation value of the energy--momentum tensor for a conformally coupled
scalar field was investigated in dS spacetime with static coordinates in
presence of curved branes, on which the field obeys the Robin boundary
conditions with coordinate dependent coefficients. In those papers the
conformal relation between dS and Rindler spacetimes and the results for the
Rindler counterpart were used. More recently, the Casimir density in a dS
spacetime with toroidally compactified spatial dimensions has been
investigated in \cite{Saha08}.

The outline of the paper is as follows. In the next section the positive
frequency Wightman function will be evaluated for a scalar field with
general curvature coupling parameter with Robin boundary conditions on two
parallel plane boundaries in the background of dS spacetime. Among the most
important quantities describing the local properties of a quantum field and
the corresponding quantum back-reaction effects are the expectation values
of the field squared and of the energy-momentum tensor. These quantities, in
the region between the plates, will be investigated in Sects.~\ref{sec:phi2}
and \ref{sec:EMT}. Simple asymptotic formulae are obtained both for small
and for large plate separations. The Casimir forces acting on the plates are
obtained in Sect.~\ref{sec:Forces}. Finally, Sect.~\ref{sec:Conc} contains a
summary of the work done together with an outlook.

\section{The Wightman function in de Sitter spacetime with two parallel
plates}

\label{sec:WF}

Consider a quantum scalar field $\varphi (x)$ on a $(D+1)$-dimensional dS
spacetime background, as coming from a cosmological theory, as described in
Sect.~1, with positive cosmological constant, $\Lambda $. The corresponding
line element can be written in planar (inflationary) coordinates, which are
most appropriate for cosmological applications:%
\begin{equation}
ds^{2}=dt^{2}-e^{2t/\alpha }\sum_{i=1}^{D}(dz^{i})^{2}.  \label{ds2deSit}
\end{equation}%
Here, the parameter $\alpha $ is related to the cosmological constant
through the expression $\alpha =D(D-1)/(2\Lambda )$. For the discussion to
follow, in addition to the synchronous time coordinate, $t$, it is very
convenient to introduce the conformal time, $\tau $, defined as $\tau
=-\alpha e^{-t/\alpha } $, $-\infty <\tau <0$. In terms of this coordinate
the metric tensor takes the conformally flat form: $g_{ik}=(\alpha /\tau
)^{2}$diag$(1,-1,\ldots ,-1) $.

The dynamics of a massive scalar field with curvature coupling parameter are
governed by the Klein--Gordon equation%
\begin{equation}
\left( \nabla _{l}\nabla ^{l}+m^{2}+\xi R\right) \varphi =0,  \label{fieldeq}
\end{equation}%
where $\nabla _{l}$ is the covariant derivative operator and $%
R=D(D+1)/\alpha ^{2}$ is the Ricci scalar for dS spacetime. The special
cases $\xi =0$ and $\xi =\xi _{D}\equiv (D-1)/4D$ correspond to minimally
and to conformally coupled fields, respectively. The importance of these two
special cases comes from the fact that, in the massless limit, the
corresponding fields mimic the behavior of gravitons and photons,
respectively. Note that non-minimal coupling is required by the
renormalizability condition for interacting theories in curved spacetime
\cite{Buch84}.

In this paper we will be interested in the study of the Casimir densities
and of the mutual forces occurring for the geometry of two infinite,
parallel plates in dS spacetime. The plates are located at $z^{D}=a_{1}$ and
$z^{D}=a_{2}$, $a_{1}<a_{2}$. As the most general set up, we assume that on
these boundaries the scalar field obeys Robin boundary conditions (BCs)
\begin{equation}
(1+\beta _{j}n^{l}\nabla _{l})\varphi (x)=[1+\beta _{j}(-1)^{j-1}\partial
_{D}]\varphi (x)=0,\quad z^{D}=a_{j},\;j=1,2,  \label{boundcond}
\end{equation}%
with constant coefficients $\beta _{j}$. For $\beta _{j}=0$ these BCs reduce
to Dirichlet ones, and for $\beta _{j}=\infty $ to Neumann BCs. The choice
of different boundary conditions on the plates corresponds, in physical
terms, to using different materials for the same. The imposition of BCs
leads to a modification of the vacuum expectation values (VEVs) for physical
quantities, as compared with those in the situation without boundaries. In
the discussion below we will assume that the quantum scalar field is
prepared in a dS invariant Bunch--Davies vacuum state \cite{Bunc78}.

Among the most important characteristics of the vacuum state are the VEVs of
the field squared and of the energy-momentum tensor. These VEVs are obtained
from the corresponding positive frequency Wightman function $W(x,x^{\prime
}) $ in the coincidence limit of the arguments. The Wightman function is
also of the essence for the consideration of the response of particle
detectors at a given state of motion (see, for instance, \cite{Birr82}).
Expanding the field operator over a complete set $\left\{ \varphi _{\sigma
}(x),\varphi _{\sigma }^{\ast }(x)\right\} $ of solutions to the classical
field equation, satisfying the boundary conditions, the positive frequency
Wightman function is best expressed as the mode-sum
\begin{equation}
W(x,x^{\prime })=\langle 0|\varphi (x)\varphi (x^{\prime })|0\rangle
=\sum_{\sigma }\varphi _{\sigma }(x)\varphi _{\sigma }^{\ast }(x^{\prime }),
\label{WF}
\end{equation}%
where the collective index $\sigma $ labels the solutions.

In the region between the plates, $a_{1}<z^{D}<a_{2}$, the eigenfunctions
realizing the Bunch--Davies vacuum state and satisfying the BC on the plate
at $z^{D}=a_{1}$, have the form
\begin{equation}
\varphi _{\sigma }(x)=C_{\sigma }\eta ^{D/2}H_{\nu }^{(1)}(\eta K)\cos
[k_{D}(z^{D}-a_{1})+\alpha _{1}(k_{D})]e^{i\mathbf{k}\cdot \mathbf{z}},
\label{eigfuncD}
\end{equation}%
with the notations $\eta =|\tau |$, $K=\sqrt{k^{2}+k_{D}^{2}}$, and%
\begin{equation}
\;e^{2i\alpha _{1}(x)}=\frac{i\beta _{1}x-1}{i\beta _{1}x+1}.\;
\label{alfa1}
\end{equation}%
In Eq.~(\ref{eigfuncD}), $\;\mathbf{z}=(z^{1},\ldots ,z^{D-1})$ is the
position vector along the dimensions parallel to the plates, $\mathbf{k}%
=(k_{1},\ldots ,k_{D-1})$, and the order of the Hankel function $H_{\nu
}^{(1)}(x)$ is given by%
\begin{equation}
\nu =\left[ D^{2}/4-D(D+1)\xi -m^{2}\alpha ^{2}\right] ^{1/2}.  \label{nu}
\end{equation}%
Note that $\nu $\ is either real and nonnegative or purely imaginary. For a
conformally coupled massless field $\nu =1/2$ and the function $H_{\nu
}^{(1)}(x)$ in (\ref{eigfuncD}) is expressed in terms of elementary
functions. From the boundary condition on the second plate $z^{D}=a_{2}$ we
find that the eigenvalues for $k_{D}$ are solutions of the transcendental
equation
\begin{equation}
(1-b_{1}b_{2}y^{2})\sin y-(b_{1}+b_{2})y\cos y=0,\;  \label{kDvalues}
\end{equation}%
where $y=k_{D}a$ and $b_{j}=\beta _{j}/a$, with $a=a_{2}-a_{1}$ being the
separation between the plates. In the discussion below we will assume that
all zeros are real. In particular, this is the case for the conditions $%
b_{j}\leqslant 0$ (see \cite{Rome02}). The positive solutions of Eq.~(\ref%
{kDvalues}) will be denoted by $y=\lambda _{n}$, $n=1,2,\ldots $, and for
the eigenvalues of $k_{D}$ one has $k_{D}=\lambda _{n}/a$. The
eigenfunctions are specified by a set of eigenfunctions $\sigma =(\mathbf{k}%
,n)$.

The coefficient $C_{\sigma }$ in (\ref{eigfuncD}) is determined from the
orthonormalization condition
\begin{equation}
\int d\mathbf{z}\int_{a_{1}}^{a_{2}}dz^{D}\sqrt{|g|}g^{00}\left[ \varphi
_{\sigma }(x)\partial _{\tau }\varphi _{\sigma ^{\prime }}^{\ast
}(x)-\varphi _{\sigma ^{\prime }}^{\ast }(x)\partial _{\tau }\varphi
_{\sigma }(x)\right] =i\delta _{nn^{\prime }}\delta (\mathbf{k-k}^{\prime }).
\label{normcond}
\end{equation}%
By using the Wronskian relation for the Hankel functions, we find%
\begin{equation}
C_{\sigma }^{2}=\frac{(2\pi )^{2-D}\alpha ^{1-D}e^{i(\nu -\nu ^{\ast })\pi
/2}}{4a\left\{ 1+\cos [\lambda _{n}+2\alpha _{1}(\lambda _{n}/a)]\sin
(\lambda _{n})/\lambda _{n}\right\} },  \label{normCD}
\end{equation}%
the star meaning complex conjugate.

Combining Eqs.~(\ref{WF}), (\ref{eigfuncD}), (\ref{normCD}), for the
Wightman function in the region between the plates, we find%
\begin{eqnarray}
W(x,x^{\prime }) &=&\frac{4(\eta \eta ^{\prime })^{D/2}}{(2\pi )^{D}a\alpha
^{D-1}}\int d\mathbf{k}\,e^{i\mathbf{k}\cdot (\mathbf{z}-\mathbf{z}^{\prime
})}\sum_{n=1}^{\infty }K_{\nu }(\eta k_{n}e^{-\pi i/2})K_{\nu }(\eta
^{\prime }k_{n}e^{\pi i/2})  \notag \\
&&\times \frac{\cos [\lambda _{n}(z^{D}-a_{1})/a+\alpha _{1}(\lambda
_{n}/a)]\cos [\lambda _{n}(z^{D\prime }-a_{1})/a+\alpha _{1}(\lambda _{n}/a)]%
}{1+\cos [\lambda _{n}+2\alpha _{1}(\lambda _{n}/a)]\sin (\lambda
_{n})/\lambda _{n}},  \label{WF1}
\end{eqnarray}%
with $k_{n}=\sqrt{k^{2}+\lambda _{n}^{2}/a^{2}}$ and where we have written
the Hankel functions in terms of the modified Bessel function $K_{\nu }(x)$.
It is well known that in dS spacetime without boundaries the Bunch--Davies
vacuum state is not a physically realizable state for {\textrm{Re\thinspace }%
}$\nu \geqslant D/2$. The corresponding Wightman function contains infrared
divergences arising from long-wavelength modes. In the presence of
boundaries, the BCs on the quantized field may exclude these modes and the
Bunch--Davies vacuum becomes a realizable state. An example of this type of
situation is provided by the geometry of two parallel plates described
above. In the region between the plates and for BCs with $\beta
_{j}\leqslant 0$, $\beta _{j}\neq \infty $, there is a maximum wavelength, $%
2\pi a/\lambda _{1}$, and the two-point function (\ref{WF1}) contains no
infrared divergences. Mathematically, this situation corresponds to the 
one where in the arguments of the modified Bessel functions we have 
$k_{n}\geqslant \lambda_{1}/a$.

As we do not know the explicit expression for $\lambda _{n}$ as a function
of $n$, and being the summand in (\ref{WF1}) a strongly oscillating function
for large values of $n$, this formula is not convenient for the evaluation
of the VEVs of the field squared and of the energy-momentum tensor. In order
to obtain a useful alternative representation, we apply to the series on $%
n $ the summation formula \cite{Rome02,Saha08b}%
\begin{eqnarray}
\sum_{n=1}^{\infty }\frac{\pi \lambda _{n}f(\lambda _{n})}{\lambda _{n}+\sin
(\lambda _{n})\cos [\lambda _{n}+2\alpha _{1}(\lambda _{n}/a)]} &=&-\frac{%
\pi }{2}\frac{f(0)}{1-b_{2}-b_{1}}+\int_{0}^{\infty }dzf(z)  \notag \\
&&+i\int_{0}^{\infty }dz\frac{f(iz)-f(-iz)}{\frac{(b_{1}z-1)(b_{2}z-1)}{%
(b_{1}z+1)(b_{2}z+1)}e^{2z}-1},  \label{sumfor}
\end{eqnarray}%
being%
\begin{eqnarray}
f(z) &=&K_{\nu }(\eta e^{-\pi i/2}\sqrt{k^{2}+z^{2}/a^{2}})K_{\nu }(\eta
^{\prime }e^{\pi i/2}\sqrt{k^{2}+z^{2}/a^{2}})  \notag \\
&&\times \cos [z(z^{D}-a_{1})/a+\alpha _{1}(z/a)]\cos [z(z^{D\prime
}-a_{1})/a+\alpha _{1}(z/a)].  \label{fz}
\end{eqnarray}%
In the case $b_{j}>0$ this function has poles on the imaginary axis and the
corresponding residue terms should be added to the right hand side of the
summation formula. In order to easy the presentation, in the discussion
below we will just consider the case $b_{j}\leqslant 0$, but a similar
procedure is valid in the general case.

Use of the summation formula (\ref{sumfor}) with (\ref{fz}) allows us to
write the Wightman function in the decomposed form%
\begin{eqnarray}
W(x,x^{\prime }) &=&W_{1}(x,x^{\prime })+\frac{2\alpha ^{1-D}}{(2\pi )^{D}}%
\int d\mathbf{k}\,e^{i\mathbf{k}\cdot (\mathbf{z}-\mathbf{z}^{\prime })}
\notag \\
&&\times \int_{k}^{\infty }du\frac{\cosh [u(z^{D}-a_{1})+\tilde{\alpha}%
_{1}(u)]\cosh [u(z^{D\prime }-a_{1})+\tilde{\alpha}_{1}(u)]}{%
c_{1}(u)c_{2}(u)e^{2au}-1}  \notag \\
&&\times y^{-D}\left[ \tilde{I}_{\nu }(\eta ^{\prime }y)\tilde{K}_{\nu
}(\eta y)+\tilde{I}_{\nu }(\eta y)\tilde{K}_{\nu }(\eta ^{\prime }y)\right]
_{y=\sqrt{u^{2}-k^{2}}},  \label{WF2}
\end{eqnarray}%
where the function $\tilde{\alpha}_{1}(u)$ is defined by the relation $e^{2%
\tilde{\alpha}_{1}(u)}=c_{1}(u)$ and we have introduced the notations%
\begin{equation}
\tilde{K}_{\nu }(y)=y^{D/2}K_{\nu }(y),\;\tilde{I}_{\nu }(y)=y^{D/2}\left[
I_{\nu }(y)+I_{-\nu }(y)\right] ,  \label{tildenot}
\end{equation}%
and%
\begin{equation}
c_{j}(u)=\frac{\beta _{j}u-1}{\beta _{j}u+1}.  \label{cju}
\end{equation}%
Note that one has $c_{j}(u)=-1$ in the case of Dirichlet BCs and $c_{j}(u)=1$
for Neumann BCs. In Eq.~(\ref{WF2}),
\begin{eqnarray}
W_{1}(x,x^{\prime }) &=&\frac{8(\eta \eta ^{\prime })^{D/2}}{(2\pi
)^{D+1}\alpha ^{D-1}}\int d\mathbf{k}\,e^{i\mathbf{k}\cdot (\mathbf{z}-%
\mathbf{z}^{\prime })}\int_{0}^{\infty }du\,\cos [u(z^{D}-a_{1})+\alpha
_{1}(u)]  \notag \\
&&\times \cos [u(z^{D\prime }-a_{1})+\alpha _{1}(u)]K_{\nu }(\eta \sqrt{%
k^{2}+u^{2}}e^{-\pi i/2})K_{\nu }(\eta ^{\prime }\sqrt{k^{2}+u^{2}}e^{\pi
i/2})  \label{WFSingle}
\end{eqnarray}%
is the Wightman function corresponding to a single plate at $z^{D}=a_{1}$.

The VEVs for the geometry of a single plate were investigated in \cite%
{Saha09}. Denoting the Wightman function for the dS spacetime without
boundaries by $W_{\text{dS}}(x,x^{\prime })$, the corresponding Wightman
function for a plate located at $z^{D}=a_{j}$ can be written in the form%
\begin{equation}
W_{j}(x,x^{\prime })=W_{\text{dS}}(x,x^{\prime })+W_{j}^{(1)}(x,x^{\prime }),
\label{Wplj}
\end{equation}%
where the part induced by the plate is given by the expression
\begin{eqnarray}
W_{j}^{(1)}(x,x^{\prime }) &=&\frac{\alpha ^{1-D}}{2(2\pi )^{D}}\int d%
\mathbf{k}\,e^{i\mathbf{k}\cdot (\mathbf{z}-\mathbf{z}^{\prime
})}\int_{k}^{\infty }du\,\frac{e^{-u|z^{D}+z^{D\prime }-2a_{j}|}}{c_{j}(u)}
\notag \\
&&\times y^{-D}\left[ \tilde{I}_{\nu }(\eta ^{\prime }y)\tilde{K}_{\nu
}(\eta y)+\tilde{I}_{\nu }(\eta y)\tilde{K}_{\nu }(\eta ^{\prime }y)\right]
_{y=\sqrt{u^{2}-k^{2}}}.  \label{WFSingle2}
\end{eqnarray}%
The two-point function in the dS spacetime without boundaries was
investigated in \cite{Bunc78,Cand75} (see also \cite{Birr82}). It is given
by the formula%
\begin{equation}
W_{\text{dS}}(x,x^{\prime })=\frac{\alpha ^{1-D}\Gamma (D/2+\nu )\Gamma
(D/2-\nu )}{2^{(D+3)/2}\pi ^{(D+1)/2}\left( u^{2}-1\right) ^{(D-1)/4}}P_{\nu
-1/2}^{(1-D)/2}(u),  \label{WFdS}
\end{equation}%
where $P_{\nu }^{\mu }(x)$ is the associated Legendre function of the first
kind and
\begin{equation}
u=-1+\frac{\sum_{l=1}^{D}(z^{l}-z^{\prime l})^{2}-(\eta -\eta ^{\prime })^{2}%
}{2\eta \eta ^{\prime }}.  \label{u}
\end{equation}

By using the definition of $\tilde{\alpha}_{1}(u)$, the Wightman function in
the region between the plates can be written in the more symmetric form:%
\begin{eqnarray}
W(x,x^{\prime }) &=&W_{\text{dS}}(x,x^{\prime
})+\sum_{j=1,2}W_{j}^{(1)}(x,x^{\prime })+\frac{\alpha ^{1-D}}{2(2\pi )^{D}}%
\int d\mathbf{k}\,e^{i\mathbf{k}\cdot (\mathbf{z}-\mathbf{z}^{\prime })}
\notag \\
&&\times \int_{k}^{\infty }du\,\frac{2\cosh [u(z^{D}-z^{D\prime
})]+\sum_{j=1,2}e^{-u|z^{D}+z^{D\prime }-2a_{j}|}/c_{j}(u)}{%
c_{1}(u)c_{2}(u)e^{2au}-1}  \notag \\
&&\times y^{-D}\left[ \tilde{I}_{\nu }(\eta ^{\prime }y)\tilde{K}_{\nu
}(\eta y)+\tilde{I}_{\nu }(\eta y)\tilde{K}_{\nu }(\eta ^{\prime }y)\right]
_{y=\sqrt{u^{2}-k^{2}}},  \label{WF3}
\end{eqnarray}%
where the last term on the rhs can be referred to as the interference part.
This integral representation of the Wightman function is valid for {\textrm{%
Re\thinspace }}$\nu <1$. As it has been shown in Ref. \cite{Ford85}, the
quantized graviton field in $D=3$ dS spacetime is equivalent to a pair of
minimally coupled massless scalar fields. Thus the representation (\ref{WF3}%
) does not apply to the graviton. In the region $z^{D}<a_{1}$ ($z^{D}>a_{2}$%
) the Wightman function coincides with the corresponding function for a
single plate located at $z^{D}=a_{1}$ ($z^{D}=a_{2}$) and is given by the
expression (\ref{Wplj}), with $j=1$ ($j=2$). The results obtained in the
present paper can be applied to a more general problem where the
cosmological constant is different in separate regions $z^{D}<a_{1}$, $%
a_{1}<z^{D}<a_{2}$, and $z^{D}>a_{2}$. In this case the plate can be
considered as a simple model of a thin domain wall separating the regions
with different dS vacua.

\section{Vacuum expectation value of the field squared}

\label{sec:phi2}

Once we have the Wightman function, we can proceed to evaluate the VEV of
the field squared, by taking the coincidence limit of the arguments. In this
limit the Wightman function is divergent and some renormalization procedure
is needed. The important point here is that for points far away from the
boundaries the divergences are the same as for the dS spacetime without
boundaries. As in our previous procedure we have already extracted, from the
Wightman function, the part $W_{\text{dS}}(x,x^{\prime })$, the
renormalization of the VEVs is just reduced to the renormalization of the
part corresponding to the geometry without boundaries, which is already done
in literature. For the further discussion of the VEVs in the coincidence
limit it is convenient to use the notations
\begin{eqnarray}
H(x,y) &=&\frac{(x^{2}-y^{2})^{(D-3)/2}}{c_{1}(x/\eta )c_{2}(x/\eta
)e^{2ax/\eta }-1},  \label{Hxy} \\
g(\beta _{j}u,yu) &=&c_{j}(u)e^{2uy}+e^{-2uy}/c_{j}(u)+2.  \label{gxy}
\end{eqnarray}%
Introducing also the function%
\begin{equation}
F_{\nu }(y)=y^{D}\left[ I_{\nu }(y)+I_{-\nu }(y)\right] K_{\nu }(y),
\label{Fnuy}
\end{equation}%
The VEV of the field squared can be expressed in the decomposed form%
\begin{eqnarray}
\langle \varphi ^{2}\rangle &=&\langle \varphi ^{2}\rangle _{\text{dS}%
}+\langle \varphi ^{2}\rangle _{j}+\frac{A_{D}}{\alpha ^{D-1}}%
\int_{0}^{\infty }dy\frac{F_{\nu }(y)}{y^{D-1}}  \notag \\
&&\times \int_{y}^{\infty }dx\,\,H(x,y)g(\beta _{j}x/\eta
,|z^{D}-a_{j}|x/\eta ),  \label{phi2}
\end{eqnarray}%
where%
\begin{equation}
A_{D}=\frac{4(4\pi )^{-(D+1)/2}}{\Gamma ((D-1)/2)}.  \label{AD}
\end{equation}
In Eq. (\ref{phi2}), $\langle \varphi ^{2}\rangle _{\text{dS}}$ is the
renormalized VEV in dS spacetime without boundaries, and the part
\begin{eqnarray}
\langle \varphi ^{2}\rangle _{j} &=&\frac{A_{D}}{\alpha ^{D-1}}%
\int_{0}^{\infty }dy\frac{F_{\nu }(y)}{y^{D-1}}\int_{y}^{\infty }dx  \notag
\\
&&\times \,(x^{2}-y^{2})^{(D-3)/2}e^{-2x|z^{D}-a_{j}|/\eta }/c_{j}(x/\eta ),
\label{phi2j}
\end{eqnarray}%
is induced by a single boundary located at $z^{D}=a_{j}$. Note that, due to
the dS invariance of the Bunch--Davies vacuum state, the VEV $\langle
\varphi ^{2}\rangle _{\text{dS}}$ does not depend on the spacetime point.
The ratio $|z^{D}-a_{j}|/\eta $ in the formulae for the VEV is the proper
distance of the observation point from the plate at $z^{D}=a_{j}$ measured
in units of the curvature scale, $\alpha $ (note that $\alpha $ is the dS
horizon size). The boundary induced VEV is a function of the combinations $%
|z^{D}-a_{j}|/\eta $ and $\beta _{j}/\eta $ only. This property follows from
the maximal symmetry of the dS spacetime.

We can write the expression for the VEV of the field squared in a more
symmetric form by using (\ref{WF3})%
\begin{equation}
\langle \varphi ^{2}\rangle =\langle \varphi ^{2}\rangle _{\text{dS}%
}+\sum_{j=1,2}\langle \varphi ^{2}\rangle _{j}+\Delta \langle \varphi
^{2}\rangle ,  \label{phi2b}
\end{equation}%
where the interference term is given by the expression%
\begin{equation}
\Delta \langle \varphi ^{2}\rangle =\frac{A_{D}}{\alpha ^{D-1}}%
\int_{0}^{\infty }dy\,y^{1-D}F_{\nu }(y)\int_{y}^{\infty
}dx\,\,H(x,y)h(x/\eta ,z^{D}),  \label{phi2Int}
\end{equation}%
with the notation%
\begin{equation}
h(x,z^{D})=2+\sum_{j=1,2}e^{-2x|z^{D}-a_{j}|}/c_{j}(x).  \label{hxz}
\end{equation}%
Note that the interference part is finite everywhere, including the points
on the plates. The surface divergences in the VEV are contained in the
single plate parts only. For points near the plates the total VEV is
dominated by these contributions. In particular, near the plate at $%
z^{D}=a_{j}$ one has $\langle \varphi ^{2}\rangle \sim (|z^{D}-a_{j}|/\eta
)^{1-D}$. The corresponding result for two parallel plates in Minkowski
spacetime is obtained from (\ref{phi2Int}) in the limit $\alpha \rightarrow
\infty $. In this limit one has $\nu \approx im\alpha $ and the modulus of
the order of the modified Bessel function is large. In addition, we have $%
\eta \rightarrow \alpha $. Details of the corresponding limiting transition
are given in Appendix, for the explicit case of the vacuum forces acting on
the plates.

For a conformally coupled massless scalar field ($\xi =\xi _{D}$, $m=0$) one
has $\nu =1/2$ and $F_{\nu }(y)=y^{D-1}$. In this case, for the interference
term we find%
\begin{equation}
\Delta \langle \varphi ^{2}\rangle =\frac{(\eta /\alpha )^{D-1}}{(4\pi
)^{D/2}\Gamma (D/2)}\int_{0}^{\infty }dx\,\frac{x^{D-2}h(x,z^{D})}{%
c_{1}(x)c_{2}(x)e^{2ax}-1}.  \label{phi2Conf}
\end{equation}%
This result (\ref{phi2Conf}) could also have been directly obtained by using
the fact that, in the special case under consideration, the problem is
conformally related to the corresponding one for Robin plates in Minkowski
spacetime \cite{Rome02,Saha08b}. From this relation, it follows that $\Delta
\langle \varphi ^{2}\rangle =a^{1-D}(\eta )\Delta \langle \varphi
^{2}\rangle ^{\text{(M)}}$, with scale factor $a(\eta )=\alpha /\eta $,
which leads to the result (\ref{phi2Conf}).

Formula (\ref{phi2Int}) is further simplified for Dirichlet and Neumann BCs.
Using the expansion $(e^{z}-1)^{-1}=\sum_{n=1}^{\infty }e^{-nz}$ and
explicitly integrating over $x$, we find
\begin{eqnarray}
\Delta \langle \varphi ^{2}\rangle ^{\text{(J)}} &=&\frac{2\alpha ^{1-D}}{%
(2\pi )^{D/2+1}}\sum_{n=1}^{\infty }\int_{0}^{\infty }dy\,y^{-1}F_{\nu
}(\eta y)  \notag \\
&&\times \bigg[2f_{D/2-1}(2nay)-\delta ^{\text{(J)}}\sum_{j=1,2}f_{D/2-1}(2y%
\left( na+|z^{D}-a_{j}|\right) )\bigg],  \label{phi2intD}
\end{eqnarray}%
where $f_{\mu }(x)=K_{\mu }(x)/x^{\mu }$, J=D,N for Dirichlet and Neumann
boundary conditions, $\delta ^{\text{(D)}}=1$, $\delta ^{\text{(N)}}=-1$.
Note that, as in the case of the Minkowski bulk, the coordinate dependent
parts in the VEV have opposite signs for Dirichlet and Neumann scalars. By
taking into account that for real values $0\leqslant \nu <1$ the function $%
F_{\nu }(\eta y)$ is non-negative and the function $f_{\mu }(x)$ is
monotonically decreasing, from (\ref{phi2intD}) we conclude that the
interference term for Dirichlet and Neumann scalars is always positive.

Now we turn to the investigation of the interference part in the VEV of the
field squared in the asymptotic regions of the ratio $a/\eta $. This ratio
is the proper distance between the plates in units of the dS curvature scale
$\alpha $. For small values of $a/\eta $ the main contribution to (\ref%
{phi2Int}) comes from large values of $y$, for which we have $F_{\nu
}(y)\approx y^{D-1}$. As a result, at leading order the interference part in
the VEV of the field squared coincides with the corresponding quantity for a
conformally coupled massless field, and is given by Eq.~(\ref{phi2Conf}). In
particular, this part is positive for Dirichlet or Neumann BCs on both
plates, and is negative for Dirichlet BC in one plate and Neumann on the
other.

For large proper distances between the plates one has $a/\eta \gg 1$. In
order to find the leading terms in the corresponding asymptotic expansion we
introduce in (\ref{phi2Int}) new integration variables $u=ax/\eta $ and $%
v=ay/\eta $. In these variables the argument of the function (\ref{Fnuy})
becomes $v\eta /a$ and in the limit under consideration it is small. By
using the asymptotic formulae for the modified Bessel functions to the
leading order, we have
\begin{equation}
F_{\nu }(y)\approx \sigma _{\nu }{\mathrm{Re}}\left[ \frac{2^{2\nu -1}\Gamma
(\nu )y^{D-2\nu }}{\Gamma (1-\nu )}\right] ,\;y\ll 1,  \label{FnuSmall}
\end{equation}%
where $\sigma _{\nu }=1$ and $\sigma _{\nu }=2$ for real and imaginary $\nu $,
respectively. By taking into account this relation, as the next step we
apply the integration formula
\begin{equation}
\int_{0}^{\infty }dv\,v^{1-2\nu }\int_{v}^{\infty
}du\,(u^{2}-v^{2})^{(D-3)/2}f(u)=\frac{\Gamma (1-\nu )\Gamma ((D-1)/2)}{%
2\Gamma ((D+1)/2-\nu )}\int_{0}^{\infty }dr\,r^{D-2\nu -1}f(r),
\label{IntFormul}
\end{equation}%
for a given function $f(u)$. Assuming also that $|\beta _{j}|/a\ll 1$, for
the remaining integral we can use the expression%
\begin{equation}
\int_{0}^{\infty }dx\frac{x^{D-2\nu -1}e^{-x|z^{D}-a_{j}|/a}}{\delta
_{1}\delta _{2}e^{x}-1}=\sum_{n=1}^{\infty }\frac{(\delta _{1}\delta
_{2})^{n}\Gamma (D-2\nu )}{(n+|z^{D}-a_{j}|/a)^{D-2\nu }},
\label{IntFormul2}
\end{equation}%
where $\delta _{j}=c_{j}(0)$. Note that $\delta _{j}=-1$ for non-Neumann BC (%
$|\beta _{j}|<\infty $) on the plate at $a_{j}$, while $\delta _{j}=1$ if
the BC is Neumann. Finally, by using the duplication formula for the gamma
function in (\ref{IntFormul2}), we find the following asymptotic behavior%
\begin{equation}
\Delta \langle \varphi ^{2}\rangle \approx \frac{\sigma _{\nu }\alpha ^{1-D}%
}{(4\pi )^{D/2+1}(a/\eta )^{D}}{\mathrm{Re}}[(2a/\eta )^{2\nu }g_{\nu
}(z^{D})],  \label{Phi2IntLargegen}
\end{equation}%
where we have defined the function
\begin{equation}
g_{\nu }(z)=\Gamma (\nu )\Gamma (D/2-\nu )\sum_{n=1}^{\infty }(\delta
_{1}\delta _{2})^{n}\bigg[2n^{2\nu -D}+\sum_{j=1,2}\delta
_{j}(n+|z-a_{j}|/a)^{2\nu -D}\bigg].  \label{gnuz}
\end{equation}

As it is seen from Eq. (\ref{Phi2IntLargegen}), for large separations of the
plates the behavior of the interference term is qualitatively different for
real and for imaginary values of the parameter $\nu $. For positive values of $%
\nu $, to leading order, we find%
\begin{equation}
\Delta \langle \varphi ^{2}\rangle \approx \frac{\alpha ^{1-D}g_{\nu }(z^{D})%
}{4\pi ^{D/2+1}(2a/\eta )^{D-2\nu }},  \label{phi2intasr}
\end{equation}%
For imaginary $\nu $, the asymptotic behavior at large distances is of the
form%
\begin{equation}
\Delta \langle \varphi ^{2}\rangle \approx \frac{|g_{\nu }(z^{D})|\cos
[2|\nu |\ln (2a/\eta )+\phi ]}{2\pi ^{D/2+1}\alpha ^{D-1}(2a/\eta )^{D}},
\label{phi2intasi}
\end{equation}%
with $g_{\nu }(z^{D})=|g_{\nu }(z^{D})|e^{i\phi }$. Hence, in this case the
decay of the interference part is oscillatory. At a given spatial point, the
dependence on the synchronous time coordinate has the form $\Delta \langle
\varphi ^{2}\rangle \sim e^{-Dt/\alpha }\cos [2|\nu |t/\alpha +\psi ]$. One
gets similar oscillations for the single plate parts at large distances from
the plate (see Ref.~\cite{Saha09}). Note that, for a given $a$, the limit
under consideration corresponds to the one for the late stages of the cosmological expansion.

\section{Vacuum expectation value of the energy-momentum tensor}

\label{sec:EMT}

For the evaluation of the VEV of the energy-momentum tensor in the region
between the plates, we use%
\begin{equation}
\langle 0|T_{ik}|0\rangle =\lim_{x^{\prime }\rightarrow x}\partial
_{i}\partial _{k}^{\prime }W(x,x^{\prime })+\left[ \left( \xi -1/4\right)
g_{ik}\nabla _{l}\nabla ^{l}-\xi \nabla _{i}\nabla _{k}-\xi R_{ik}\right]
\langle \varphi ^{2}\rangle ,  \label{emtvev1}
\end{equation}%
where $R_{ik}=Dg_{ik}/\alpha ^{2}$ is the Ricci tensor for the dS spacetime.
Taking advantage of the expressions for the Wightman function and for the
VEV of the field squared from the previous sections, the renormalized VEVs
for the diagonal components of the energy-momentum tensor can be expressed
in the form (no summation over $l$)
\begin{eqnarray}
\langle T_{l}^{l}\rangle &=&\langle T_{l}^{l}\rangle _{\text{dS}}+\langle
T_{l}^{l}\rangle _{j}+\frac{A_{D}}{\alpha ^{D+1}}\int_{0}^{\infty
}dy\,y^{1-D}\int_{y}^{\infty }dx\,H(x,y)  \notag \\
&&\times \left[ g(\beta _{j}x/\eta ,|z^{D}-a_{j}|x/\eta
)G_{l}(y)+2G_{l}x^{2}F_{\nu }(y)\right] ,  \label{Tll}
\end{eqnarray}%
where we have introduced the notations%
\begin{eqnarray}
G_{0}(y) &=&\left[ (y^{2}/4)\partial _{y}^{2}-D(\xi +\xi _{D})y\partial
_{y}+D^{2}\xi \right.  \notag \\
&&\left. +m^{2}\alpha ^{2}-y^{2}+\left( 1-4\xi \right) x^{2}\right] F_{\nu
}(y),  \notag \\
G_{D}(y) &=&\left\{ \left( \xi -\frac{1}{4}\right) y^{2}\partial _{y}^{2}+%
\left[ \xi (2-D)+\frac{D-1}{4}\right] y\partial _{y}-\xi D\right\} F_{\nu
}(y),  \label{Fhat} \\
G_{l}(y) &=&G_{D}(y)+\left[ \frac{y^{2}-x^{2}}{D-1}+\left( 1-4\xi \right)
x^{2}\right] F_{\nu }(y),\;l=1,\ldots ,D-1,  \notag
\end{eqnarray}%
and%
\begin{equation}
G_{D}=1,\;G_{l}=4\xi -1,\;l=0,1,\ldots ,D-1.  \label{Glln}
\end{equation}%
Note that, though not explicitly written, the functions $G_{l}(y)$ with $%
l=0,\ldots ,D-1$ do depend on $x$ as well. In Eq.~(\ref{Tll}), $\langle
T_{l}^{l}\rangle _{\text{dS}}$ is the corresponding renormalized VEV in dS
spacetime without boundaries. For points away from the plates,
renormalization is strictly necessary for this part only. Owing to the dS
invariance of the Bunch--Davies vacuum, the part $\langle T_{k}^{l}\rangle _{%
\text{dS}}$ is proportional to the metric tensor with a constant coefficient
and has been well investigated in the literature \cite{Bunc78,Cand75}. For
the part induced by a single plate at $z^{D}=a_{j}$, one has \cite{Saha09}
(no summation over $l$)%
\begin{eqnarray}
\langle T_{l}^{l}\rangle _{j} &=&\frac{A_{D}}{\alpha ^{D+1}}\int_{0}^{\infty
}dy\,y^{1-D}\int_{y}^{\infty }dx\,(x^{2}-y^{2})^{(D-3)/2}  \notag \\
&&\times \frac{e^{-2x|z^{D}-a_{j}|/\eta }}{c_{j}(x/\eta )}G_{l}(y).
\label{Tllj}
\end{eqnarray}%
The last term on the rhs of Eq.~(\ref{Tll}) is induced by the presence of
the second plate.

For the non-zero off-diagonal component, we have%
\begin{eqnarray}
\langle T_{0}^{D}\rangle &=&\langle T_{0}^{D}\rangle _{j}-\text{sgn}%
(z^{D}-a_{j})\frac{A_{D}}{2\alpha ^{D+1}}\int_{0}^{\infty }dyy^{1-D}G_{0D}(y)
\notag \\
&&\times \int_{y}^{\infty }dx\,xH(x,y)\left[ c_{j}(x/\eta
)e^{2x|z^{D}-a_{j}|/\eta }-e^{-2x|z^{D}-a_{j}|/\eta }/c_{j}(x/\eta )\right] ,
\label{T0Dn}
\end{eqnarray}%
where the part corresponding to the geometry of a single plate is given by%
\begin{eqnarray}
\langle T_{0}^{D}\rangle _{j} &=&\text{sgn}(z^{D}-a_{j})\frac{2A_{D}}{\alpha
^{D+1}}\int_{0}^{\infty }dyy^{1-D}G_{0D}(y)  \notag \\
&&\times \int_{y}^{\infty }dx\,x(x^{2}-y^{2})^{(D-3)/2}\frac{%
e^{-2x|z^{D}-a_{j}|/\eta }}{c_{j}(x/\eta )}.  \label{T0Dj}
\end{eqnarray}%
In these formulae we have defined the function%
\begin{equation}
G_{0D}(y)=\left[ (4\xi -1)y\partial _{y}+4\xi \right] F_{\nu }(y).
\label{F0D}
\end{equation}%
The off-diagonal component (\ref{T0Dn}) corresponds to the energy flux along
the direction perpendicular to the plates. This type of the energy flux also
appears in the geometry of a cosmic string on backgrounds of
Friedmann--Robertson--Walker and dS spacetimes \cite{Davi88}. Depending on
the values of the coefficients in the boundary conditions and of the field
mass this flux can be positive or negative. As an additional check of the
expressions for the energy-momentum tensor, it can be seen that the ones for
the single plate contribution and for the second plate induced part fulfill
the trace relation%
\begin{equation}
\langle T_{l}^{l}\rangle =\left[ D(\xi -\xi _{D})\nabla _{l}\nabla ^{l}+m^{2}%
\right] \langle \varphi ^{2}\rangle .  \label{tracerel}
\end{equation}%
The boundary induced part in the VEV of the energy-momentum tensor is
traceless for a conformally coupled massless scalar. The trace anomaly is
contained in the boundary-free part only.

For a conformally coupled massless scalar field ($\xi =\xi _{D}$, $m=0$) the
single plate part in the VEV of the energy-momentum tensor vanishes and one
finds
\begin{eqnarray}
\langle T_{k}^{l}\rangle &=&\langle T_{k}^{l}\rangle _{\text{dS}}-\frac{%
(\eta /\alpha )^{D+1}}{(4\pi )^{D/2}\Gamma (D/2+1)}\ \text{diag}(1,\ldots
,1,-D)  \notag \\
&&\times \int_{0}^{\infty }dx\,\frac{x^{D}}{c_{1}(x)c_{2}(x)e^{2ax}-1}\,.
\label{TllConf}
\end{eqnarray}%
As in the case of the field squared, the boundary induced part in this
formula could have been obtained from the corresponding result for the
Casimir effect in Minkowski spacetime, by using the fact that the two
problems are conformally related. The electromagnetic field in $D=3$ is
conformally invariant and the Casimir problem with two perfectly conducting
parallel plates is reduced to the corresponding problem with two scalar
modes with Dirichlet and Neumann BCs. In this case, the single plate parts
vanish and for the interference part we have $\Delta \langle
T_{l}^{k}\rangle =-(\pi ^{2}/720)(\alpha a/\eta )^{-4}$diag$(1,1,1,-3)$. In
the case $D>3$, the electromagnetic field is not conformally invariant and
we expect that the corresponding VEV will depend on the distance from the
plates. However, this case requires further consideration.

The VEVs for the components of the energy-momentum tensor can be written in
the more symmetric form%
\begin{equation}
\langle T_{k}^{l}\rangle =\langle T_{l}^{l}\rangle _{\text{dS}%
}+\sum_{j=1,2}\langle T_{k}^{l}\rangle _{j}+\Delta \langle T_{k}^{l}\rangle ,
\label{TlkDecomp}
\end{equation}%
where for the interference terms we have (no summation over $l$)%
\begin{eqnarray}
\Delta \langle T_{l}^{l}\rangle &=&\frac{A_{D}}{\alpha ^{D+1}}%
\int_{0}^{\infty }dy\,y^{1-D}\int_{y}^{\infty }dx\,H(x,y)  \notag \\
&&\times \left[ h(x/\eta ,z^{D})G_{l}(y)+2G_{l}x^{2}F_{\nu }(y)\right] ,
\label{Tll2} \\
\Delta \langle T_{0}^{D}\rangle &=&\frac{A_{D}}{2\alpha ^{D+1}}%
\int_{0}^{\infty }dyy^{1-D}\int_{y}^{\infty }dx\,xH(x,y)  \notag \\
&&\times \sum_{j=1,2}\text{sgn}(z^{D}-a_{j})\frac{e^{-2x|z^{D}-a_{j}|/\eta }%
}{c_{j}(x/\eta )}G_{0D}(y).  \label{T0D2}
\end{eqnarray}%
Note that when the coefficients in the BCs are the same, $\beta _{1}=\beta
_{2}$, the energy flux vanishes at the point $z^{D}=(a_{1}+a_{2})/2$. Of
course, this is a direct consequence of the symmetry of the problem.

In the special cases of Dirichlet and Neumann BCs, expressions similar to (%
\ref{phi2intD}) can be obtained for the interference terms in the VEVs of
the energy-momentum tensor. Here we give the corresponding formulas for the
interference parts in the normal stress and in the off-diagonal component:%
\begin{eqnarray}
\Delta \langle T_{D}^{D}\rangle &=&\frac{4\alpha ^{-D-1}}{(2\pi )^{D/2+1}}%
\sum_{n=1}^{\infty }\int_{0}^{\infty }dy\bigg\{y\left[
(D-1)f_{D/2}(z)+f_{D/2-1}(z)\right] F_{\nu }(y)+  \notag \\
&&\times \bigg[f_{D/2-1}(z)-\frac{\delta ^{\text{(J)}}}{2}%
\sum_{j=1,2}f_{D/2-1}(z+2y|z^{D}-a_{j}|/\eta )\bigg]\frac{G_{D}(y)}{y}\bigg\}%
_{z=2yna/\eta },  \label{TDDDN} \\
\Delta \langle T_{0}^{D}\rangle &=&-\frac{\delta ^{\text{(J)}}\alpha ^{-D-1}%
}{(2\pi )^{D/2+1}}\sum_{n=1}^{\infty }\int_{0}^{\infty }dy\sum_{j=1,2}\text{%
sgn}(z^{D}-a_{j})  \notag \\
&&\times zf_{D/2}(z)|_{z=2y\left( an+|z^{D}-a_{j}|\right) /\eta }G_{0D}(y),
\label{T0DDN}
\end{eqnarray}%
with J=D,N and being the function $f_{\mu }(z)$ defined after the formula (%
\ref{phi2intD}). As for the case of the field squared, the coordinate
dependent parts have opposite signs for Dirichlet and Neumann boundary
conditions, respectively.

We now wish to examine the behavior of the VEV of the energy-momentum
tensor in the asymptotic regimes of small and of large separations between the
plates. At small separation, assuming that $a/\eta \ll 1$, we introduce in (%
\ref{Tll2}), (\ref{T0D2}) new integration variables $u=ax/\eta $ and $%
v=ay/\eta $. The arguments of the functions $G_{l}(y)$ and $F_{\nu }(y)$ are
large and we can use the corresponding asymptotic formulae for the modified
Bessel functions. In particular, one has $F_{\nu }(y)\approx y^{D-1}$ and
\begin{eqnarray}
G_{0}(y) &\approx &\left[ \left( 1-4\xi \right) x^{2}-y^{2}\right]
y^{D-1},\;G_{D}(y)\approx D(\xi _{D}-\xi )y^{D-1},  \notag \\
G_{l}(y) &\approx &G_{0}(y)+\frac{Dy^{2}-x^{2}}{D-1}y^{D-1},\;l=1,\ldots
,D-1.  \label{GlSmall}
\end{eqnarray}
For the further evaluation of the integrals, we use Eq.~(\ref{IntFormul})
with $\nu =\pm 1/2$. As a result, to leading order, one has (no summation
over $l$)%
\begin{eqnarray}
\Delta \langle T_{l}^{l}\rangle &\approx &\frac{2(4\pi )^{-D/2}}{\Gamma
(D/2)(\alpha /\eta )^{D+1}}\int_{0}^{\infty }dx\frac{x^{D}[\mathcal{B}%
_{l}h(x,z^{D})+\mathcal{C}_{l}]}{c_{1}(x)c_{2}(x)e^{2ax}-1},  \notag \\
\Delta \langle T_{0}^{D}\rangle &\approx &\frac{2(4\pi )^{-D/2}D(\xi -\xi
_{D})}{\Gamma (D/2)\alpha (\alpha /\eta )^{D}}\sum_{j=1,2}\int_{0}^{\infty
}dx\frac{\text{sgn}(z^{D}-a_{j})x^{D-1}}{c_{1}(x)c_{2}(x)e^{2ax}-1}\frac{%
e^{-2x|z^{D}-a_{j}|}}{c_{j}(x)},  \label{Tllsmall}
\end{eqnarray}%
where%
\begin{equation}
\mathcal{B}_{l}=-2\left( \xi -\xi _{D}\right) ,\;\mathcal{B}_{D}=0,\;%
\mathcal{C}_{l}=4\xi -1,\;\mathcal{C}_{D}=1,\;  \label{BlCl}
\end{equation}%
with $l=0,1,\ldots ,D-1$. Note that $\Delta \langle T_{0}^{D}\rangle \sim
(a/\eta )\Delta \langle T_{l}^{l}\rangle $ and that, for a conformally
coupled field, the leading terms in the diagonal components are homogeneous.
For small separation, to leading order, the energy density is equal to the
stresses along the directions parallel to the plates. The vacuum stress
normal to the plates, $\Delta \langle T_{D}^{D}\rangle $, is positive for
Dirichlet and Neumann BCs, and is negative for Dirichlet BC on one plate and
Neumann on the other.

Now let us discuss the asymptotics at large distances between the plates, $%
a/\eta \gg 1$, when the curvature effects are essential. The corresponding
asymptotic formulae for the the interference parts in the VEV of the
energy-momentum tensor are found in a way similar to that already described
for the VEV of the field squared in Sect. \ref{sec:phi2}. For positive
values of the parameter $\nu $, the leading terms have the form (no
summation over $l$)%
\begin{eqnarray}
\Delta \langle T_{l}^{l}\rangle &\approx &\frac{f_{l}\alpha ^{-D-1}g_{\nu
}(z^{D})}{4\pi ^{D/2+1}\left( 2a/\eta \right) ^{D-2\nu }},  \notag \\
\Delta \langle T_{0}^{D}\rangle &\approx &\frac{\alpha ^{-D-1}g_{\nu
}^{(0)}(z^{D})}{4\pi ^{D/2+1}(2a/\eta )^{D-2\nu +1}},  \label{TllLarge}
\end{eqnarray}%
where the function $g_{\nu }(z)$ is defined by Eq.~(\ref{gnuz}) and
\begin{eqnarray}
g_{\nu }^{(0)}(z) &=&\left[ (4\xi -1)(D-2\nu )+4\xi \right] \Gamma (D/2-\nu
+1)  \notag \\
&&\times \Gamma (\nu )\sum_{n=1}^{\infty }\sum_{j=1,2}\frac{\text{sgn}%
(z-a_{j})(\delta _{1}\delta _{2})^{n}\delta _{j}}{(n+|z-a_{j}|/a)^{D-2\nu +1}%
}.  \label{gnu0}
\end{eqnarray}%
The coefficients $f_{l}$ for separate components of the energy-momentum
tensor are defined as%
\begin{equation}
f_{l}=(2\nu /D)f_{0}=-2\nu \left[ \xi +\left( \xi -1/4\right) (D-2\nu )%
\right] ,  \label{fl}
\end{equation}%
with $l=1,\ldots ,D$. Note that for minimally and conformally coupled
massless fields $f_{0}=f_{l}=0$ and thus the leading terms vanish. As we
see, in the limit under consideration the vacuum stresses are isotropic and $%
|\Delta \langle T_{0}^{D}\rangle |\ll |\Delta \langle T_{l}^{l}\rangle |$.
The corresponding equation of state is of barotropic type: $\Delta \langle
T_{1}^{1}\rangle \approx \cdots \approx \Delta \langle T_{D}^{D}\rangle
\approx (2\nu /D)\Delta \langle T_{0}^{0}\rangle $. A similar relation holds
for the single plate parts at large distances \cite{Saha09}. As a
consequence, the equation of state parameter for the total boundary-induced
part is equal to $-2\nu /D$, and it is negative.

For imaginary $\nu $ the leading terms at large distances are given by%
\begin{eqnarray}
\Delta \langle T_{l}^{l}\rangle &\approx &\frac{\alpha ^{-D-1}|g_{\nu
}(z^{D})||f_{l}|}{2\pi ^{D/2+1}\left( 2a/\eta \right) ^{D}}\cos [2|\nu |\ln
\left( 2a/\eta \right) +\phi +\phi _{l}],  \notag \\
\Delta \langle T_{0}^{D}\rangle &\approx &\frac{\alpha ^{-D-1}|g_{\nu
}^{(0)}(z^{D})|}{2\pi ^{D/2+1}(2a/\eta )^{D+1}}\cos [2|\nu |\ln \left(
2a/\eta \right) +\phi _{0}^{(D)}].  \label{TllLargeIm}
\end{eqnarray}%
In these formulae $f_{l}=|f_{l}|e^{i\phi _{l}}$ and $g_{\nu
}^{(0)}(z^{D})=|g_{\nu }^{(0)}(z^{D})|e^{i\phi _{0}^{(D)}}$. In this case,
the damping of the interference parts as functions of the proper distance is
oscillatory. In terms of the synchronous time coordinate, at a given spatial
point we have damping oscillations in accordance with $\Delta \langle
T_{l}^{l}\rangle \sim e^{-Dt/\alpha }\cos (2|\nu |t/\alpha +\psi _{l})$.
From (\ref{fl}) it follows that the oscillations in the energy density and
in the vacuum stresses are shifted in phase by $\pi /2$.

\begin{figure}[tbph]
\begin{center}
\begin{tabular}{cc}
\epsfig{figure=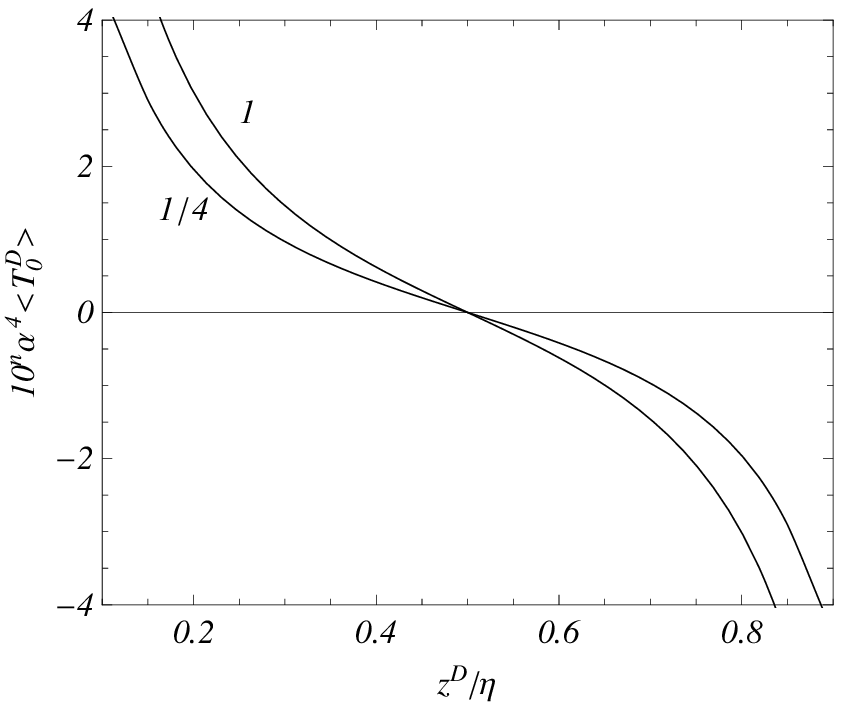,width=7.cm,height=6.cm} & \quad %
\epsfig{figure=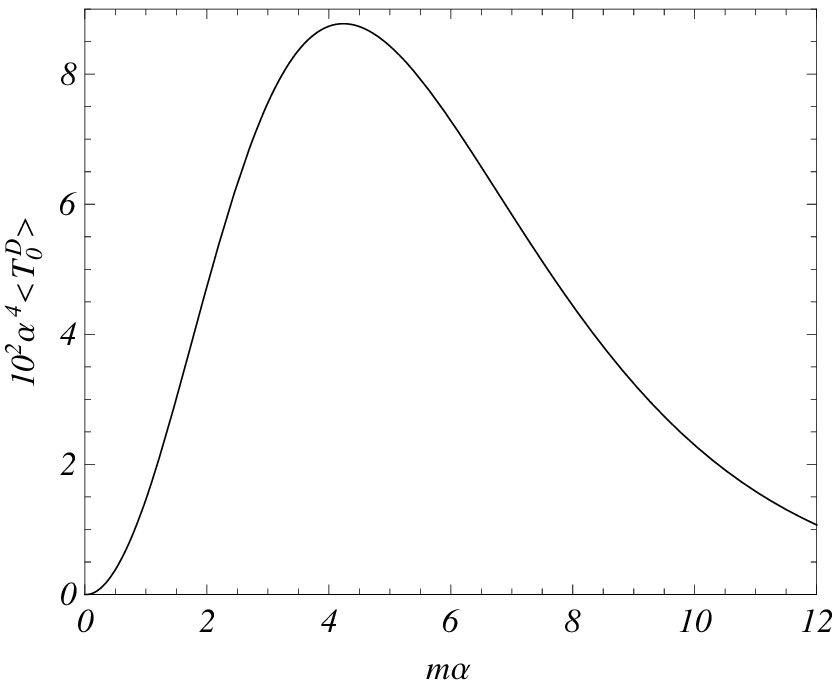,width=7.cm,height=6.cm}%
\end{tabular}%
\end{center}
\caption{Energy flux in the region between the plates as a function of $%
z^{D}/\protect\eta $ for $a/\protect\eta =1$, $m\protect\alpha =1/4,1$ (left
plot), and as a function of $m\protect\alpha $ for $a/\protect\eta =1$, $%
z^{D}/\protect\eta =0.3$ (right plot), for a $D=3$ conformally coupled
scalar field with Dirichlet boundary condition. On the left plot the figures
near the curves are the values of the parameter $m\protect\alpha $, and the
scaling factor $n$ is equal to 2 and 3, for $m\protect\alpha =1$ and $m%
\protect\alpha =1/4$, respectively.}
\label{fig1}
\end{figure}

In Fig.~\ref{fig1}, for the case of a $D=3$ conformally coupled scalar field
with Dirichlet BCs on both plates, we have plotted the energy flux as a
function of $z^{D}/\eta $, for given values of $a/\eta =1$ and $m\alpha
=1/4,1$ (left plot), and as a function of $m\alpha $ for given values of $%
a/\eta =1$ and $z^{D}/\eta =0.3$ (right plot). For Neumann BCs the flux has
the opposite sign.

\section{Casimir force}

\label{sec:Forces}

In this section we consider the vacuum forces acting on the plates. The
vacuum force acting per unit surface of the plate at $z^{D}=a_{j}$ is
determined by the $_{D}^{D}$-component of the vacuum energy-momentum tensor
evaluated at this point. For the region between the plates, the
corresponding effective pressures can be written as a sum of two terms,
namely%
\begin{equation}
p^{(j)}=p_{1}^{(j)}+p_{\text{(int)}}^{(j)},\;j=1,2.  \label{pj}
\end{equation}%
The first term on the rhs is the pressure for a single plate at $z^{D}=a_{j}$%
, when the second plate is absent. This term is divergent due to the surface
divergences in the subtracted vacuum expectation values and needs additional
renormalization. The second term on the rhs of Eq.~(\ref{pj}) is the
pressure induced by the presence of the second plate, and can be termed as
an interaction force. This contribution is finite for all nonzero distances
between the plates. In the regions $z^{D}<a_{1}$ and $z^{D}>a_{2}$ we have $%
p^{(j)}=p_{1}^{(j)}$. As a result, the contributions to the vacuum force
coming from the term $p_{1}^{(j)}$ are the same from the left and from the
right sides of the plate, so that there is no net contribution to the
effective force.

The interaction force on the plate at $z^{D}=a_{j}$ is obtained from the
last term on the rhs of Eq.~(\ref{Tll}) for $\langle T_{D}^{D}\rangle $
(with minus sign) substituting $z^{D}=a_{j}$, that is%
\begin{eqnarray}
p_{\text{(int)}}^{(j)} &=&-\frac{2A_{D}}{\alpha ^{D+1}}\int_{0}^{\infty
}dy\,y^{1-D}\int_{y}^{\infty }dx\,x^{2}H(x,y)  \notag \\
&&\times \left[ \frac{2\left( \beta _{j}/\eta \right) ^{2}G_{D}(y)}{\left(
\beta _{j}x/\eta \right) ^{2}-1}+F_{\nu }(y)\right] .  \label{pintj}
\end{eqnarray}%
Depending on the values of the coefficients in the boundary conditions, the
effective pressures (\ref{pintj}) can be either positive or negative,
leading to repulsive or to attractive forces, respectively. For $\beta
_{1}\neq \beta _{2}$ the Casimir forces acting on the left and on the right
plates are different. In the Appendix we show that, as it must be, in the
limit $\alpha \rightarrow \infty $ the corresponding result for the geometry
of two parallel plates in Minkowski spacetime is obtained.

The general formula is further simplified for the special cases of Dirichlet
and of Neumann boundary conditions. For Dirichlet BCs on both plates, we find%
\begin{equation}
p_{\text{(int)}}^{(\text{D})}=-\frac{4\alpha ^{-D-1}}{(2\pi )^{D/2+1}}%
\sum_{n=1}^{\infty }\int_{0}^{\infty }dy\,yF_{\nu }(y)\left[
(D-1)f_{D/2}(2nay/\eta )+f_{D/2-1}(2nay/\eta )\right] .  \label{pjD}
\end{equation}%
For $0\leqslant \nu <1$ the integrand in this formula is positive and $p_{%
\text{(int)}}^{(\text{D})}$ is negative, yielding an attractive force for
all separations. In the case of Neumann BCs the corresponding expression is
\begin{equation}
p_{\text{(int)}}^{(\text{N})}=p_{\text{(int)}}^{(\text{D})}-\frac{8\alpha
^{-D-1}}{(2\pi )^{D/2+1}}\sum_{n=1}^{\infty }\int_{0}^{\infty }dy\,\frac{%
G_{D}(y)}{y}f_{D/2-1}(2nay/\eta ),  \label{pjN}
\end{equation}%
where the function $G_{D}(y)$ is defined in Eq.~(\ref{Fhat}).

Let us now investigate the asymptotic behavior of the vacuum forces. This
can be done in the way similar to that already described in the case of the
VEV for the energy-momentum tensor. In the limit of small proper distances
between the plates, $a/\eta \ll 1$, the main contribution to the integral
over $y$ comes from large values of $y$, $y\sim \eta /a$. By using the
corresponding asymptotics for the modified Bessel functions, to leading
order we find%
\begin{equation}
p_{\text{(int)}}^{(j)}\approx -\frac{2(\eta /\alpha )^{D+1}}{(4\pi
)^{D/2}\Gamma (D/2)}\int_{0}^{\infty }dx\,\frac{x^{D}}{%
c_{1}(x)c_{2}(x)e^{2ax}-1}.  \label{pintjsmall}
\end{equation}%
If, in addition, $|\beta _{j}|/a\gg 1$, one has%
\begin{equation}
p_{\text{(int)}}^{(j)}\approx -\frac{D\Gamma ((D+1)/2)\zeta _{\text{R}}(D+1)%
}{(4\pi )^{(D+1)/2}(\alpha a/\eta )^{D+1}},  \label{pintjsmall2}
\end{equation}%
and the corresponding force is attractive. In (\ref{pintjsmall2}), $\zeta _{%
\text{R}}(x)$ is the Riemann zeta function. The same result is obtained for
Dirichlet BCs on both plates. In the case of Dirichlet BC on one plate and
non-Dirichlet one on the other, the leading term is obtained from (\ref%
{pintjsmall2}) with an additional factor $(2^{-D}-1)$. In this case the
vacuum force at small distance is repulsive.

Now we consider the large distance asymptotics, $a/\eta \gg 1$. The cases of
real and imaginary $\nu $ must be studied separately. For positive values of
$\nu $, one has%
\begin{eqnarray}
p_{\text{(int)}}^{(j)} &\approx &-\frac{2\alpha ^{-D-1}g_{\nu }^{(j)}}{\pi
^{D/2+1}(2a/\eta )^{D-2\nu +2}},\;|\beta _{j}|<\infty ,  \notag \\
p_{\text{(int)}}^{(j)} &\approx &-\frac{\alpha ^{-D-1}g_{\nu }^{\text{N}(j)}%
}{\pi ^{D/2+1}(2a/\eta )^{D-2\nu }},\;\beta _{j}=\infty ,  \label{pjlarge}
\end{eqnarray}%
where we have introduced the notations%
\begin{eqnarray}
g_{\nu }^{(j)} &=&\left[ (D+1)/2-\nu \right] \Gamma (D/2-\nu +1)\Gamma (\nu )
\notag \\
&&\times \left[ 1-2(\beta _{j}/\eta )^{2}f_{D}\right] \sum_{n=1}^{\infty }%
\frac{(\delta _{1}\delta _{2})^{n}}{n^{D-2\nu +2}},  \label{gnuj} \\
g_{\nu }^{\text{N}(j)} &=&\Gamma (D/2-\nu )\Gamma (\nu
)f_{D}\sum_{n=1}^{\infty }\frac{(\delta _{1}\delta _{2})^{n}}{n^{D-2\nu }},
\notag
\end{eqnarray}%
and $f_{D}$ is defined by Eq.~(\ref{fl}). Note that $\beta _{j}=\infty $
corresponds to Neumann BC. In the case of non-Neumann BCs we have assumed
that $|\beta _{j}|/a\ll 1$. As it is seen from (\ref{pjlarge}), when $%
f_{D}\neq 0$, at large distances the ratio of the Casimir forces acting on
the plate with Neumann and non-Neumann BCs is of the order $(a/\eta )^{2}$.
Note that in neither of these cases does the force depend on the specific
value of Robin coefficient in the BC on the second plate. For Dirichlet BC
on the plate at $z^{D}=a_{j}$ ($\beta _{j}=0$), at large separations the
Casimir force acting on that plate is repulsive (attractive) for Neumann
(non-Neumann) BCs on the other plate. The nature of the force acting on the
plate with Neumann BC depends on the sign of $f_{D}$ and can be either
repulsive or attractive, in function of the curvature coupling parameter and
of the field mass. For minimally and conformally coupled massive scalar
fields one has $f_{D}=\nu (D/2-\nu )$ and $f_{D}=\nu (1/2-\nu )/D$,
respectively, and this parameter is positive. The corresponding force is
attractive (repulsive) for Neumann (non-Neumann) BC on the second plate.
Note that for the geometry of parallel plates in the Minkowski bulk the
Casimir forces at large distances are repulsive for Neumann BC on one plate
and for non-Neumann BC on the other plate. For all other cases of BCs the
forces are attractive.

At large separations between the plates and for imaginary $\nu $, the
leading order terms in the corresponding asymptotic expansions take the form%
\begin{eqnarray}
p_{\text{(int)}}^{(j)} &\approx &-\frac{4\alpha ^{-D-1}|g_{\nu }^{(j)}|}{\pi
^{D/2+1}(2a/\eta )^{D+2}}\cos [2|\nu |\ln (2a/\eta )+\phi _{(j)}],\;|\beta
_{j}|<\infty ,  \notag \\
p_{\text{(int)}}^{(j)} &\approx &-\frac{2\alpha ^{-D-1}|g_{\nu }^{\text{N}%
(j)}|}{\pi ^{D/2+1}(2a/\eta )^{D}}\cos [2|\nu |\ln (2a/\eta )+\phi _{(j)}^{%
\text{N}}],\;\beta _{j}=\infty ,  \label{pjlargeim}
\end{eqnarray}%
where we have defined $g_{\nu }^{(j)}=|g_{\nu }^{(j)}|e^{i\phi _{(j)}}$ and $%
g_{\nu }^{\text{N}(j)}=|g_{\nu }^{\text{N}(j)}|e^{i\phi _{(j)}^{\text{N}}}$.
In such case the decay of the vacuum forces is oscillatory. In terms of the
synchronous time coordinate one gets the behavior $p_{\text{(int)}%
}^{(j)}\sim \exp [(D+2-2\delta _{0,1/\beta _{j}})t/\alpha ]\cos [2|\nu
|t/\alpha +\psi _{p}]$.

\begin{figure}[tbph]
\begin{center}
\begin{tabular}{cc}
\epsfig{figure=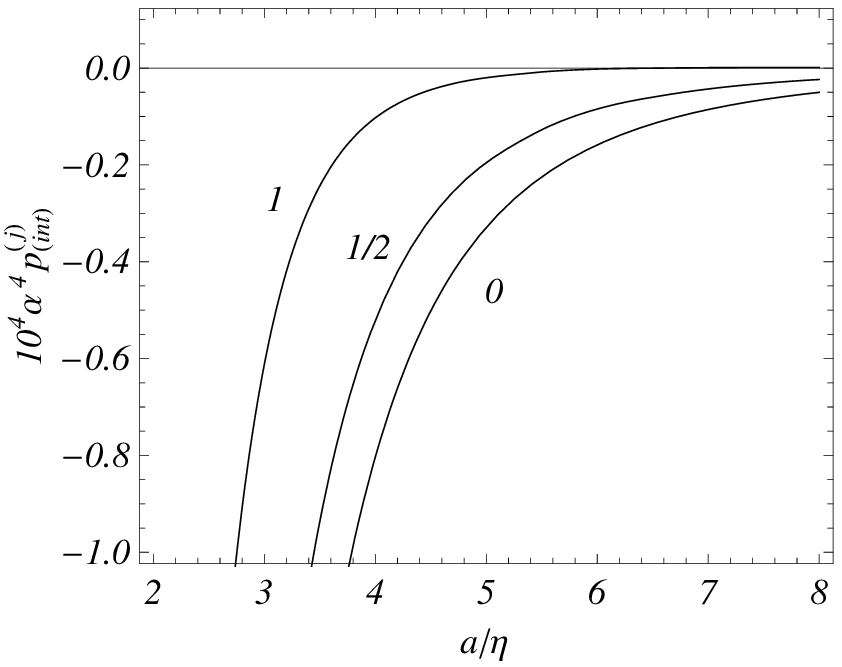,width=7.cm,height=6.cm} & \quad %
\epsfig{figure=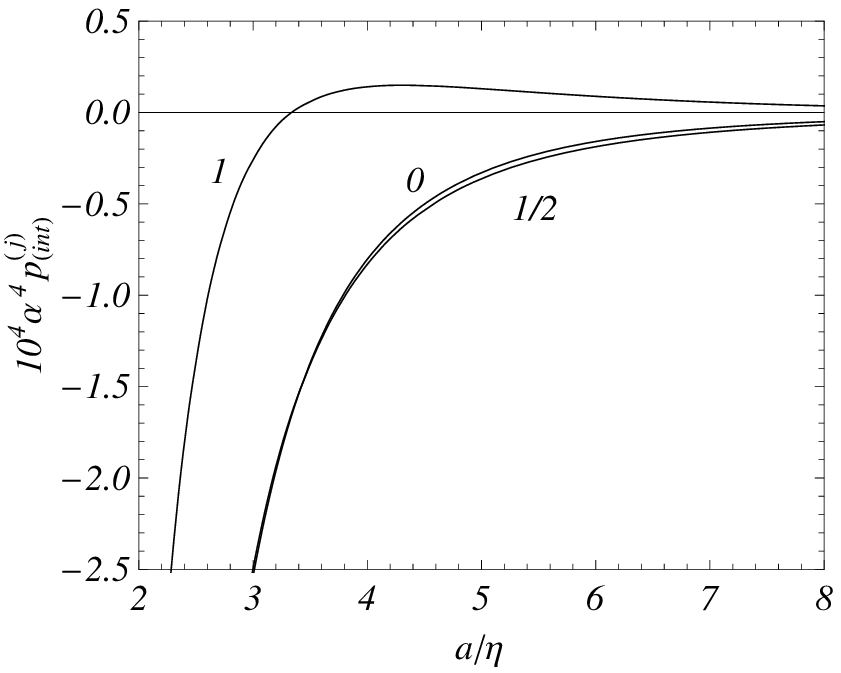,width=7.cm,height=6.cm}%
\end{tabular}%
\end{center}
\caption{Interaction forces between the plates for a $D=3$ conformally
coupled scalar field with Dirichlet (left plot) and Neumann (right plot)
BCs. The figures near the curves are the values of the parameter $\protect%
\alpha m$. Note in particular the plots corresponding to a massless field in the two cases.}
\label{fig2n}
\end{figure}

\begin{figure}[tbph]
\begin{center}
\begin{tabular}{cc}
\epsfig{figure=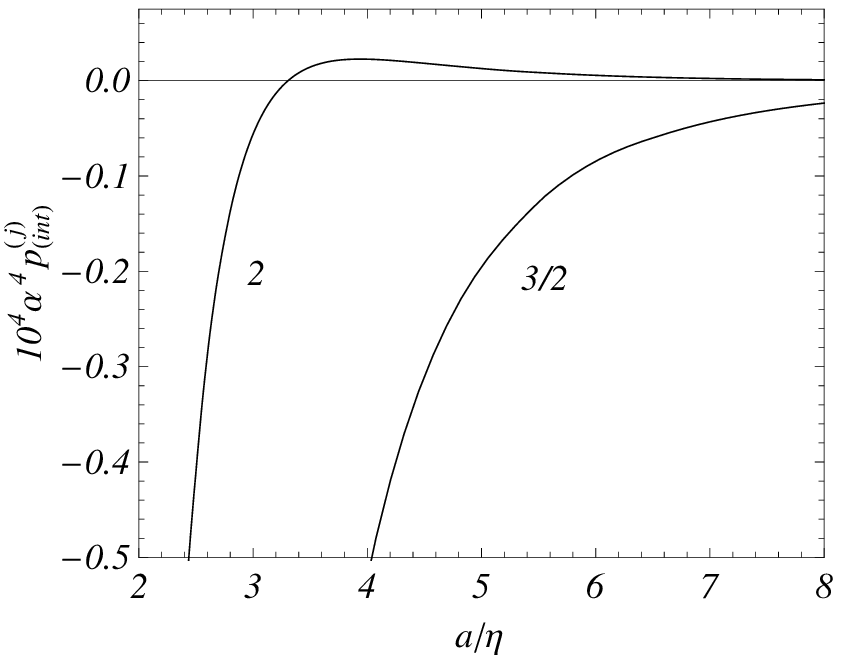,width=7.cm,height=6.cm} & \quad %
\epsfig{figure=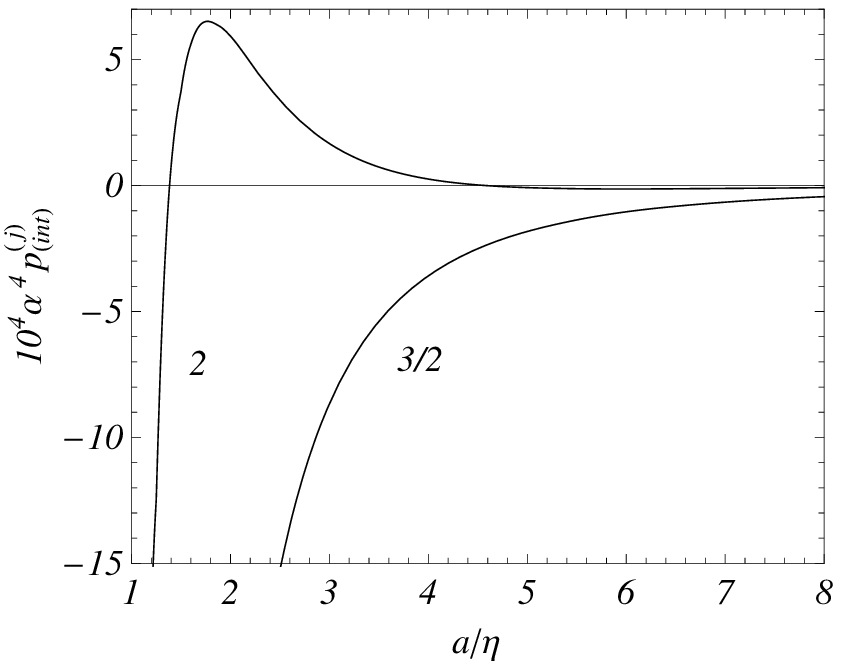,width=7.cm,height=6.cm}%
\end{tabular}%
\end{center}
\caption{Same as in Fig.~\protect\ref{fig2n} for a minimally coupled scalar
field. }
\label{fig3n}
\end{figure}

In Figs.~\ref{fig2n} and \ref{fig3n} we have plotted the Casimir force for a
$D=3$ scalar field, conformally and minimally coupled to gravity,
respectively, as a function of the proper distance between the plates,
measured in units of the dS curvature scale, $\alpha $. The left
(resp.~right) plots are for Dirichlet (resp.~Neumann) BCs on both plates.
The figures near the curves correspond to the values of the parameter $%
m\alpha $.  Values are taken in a way so to have both possibilities,
with positive and purely imaginary values of the parameter $\nu $, with
corresponding monotonic and oscillatory behavior of the forces at large
distances, respectively. Note also, in particular, the plots corresponding to 
a massless field in the two cases, corresponding to a photon-like contribution.

\begin{figure}[tbph]
\begin{center}
\epsfig{figure=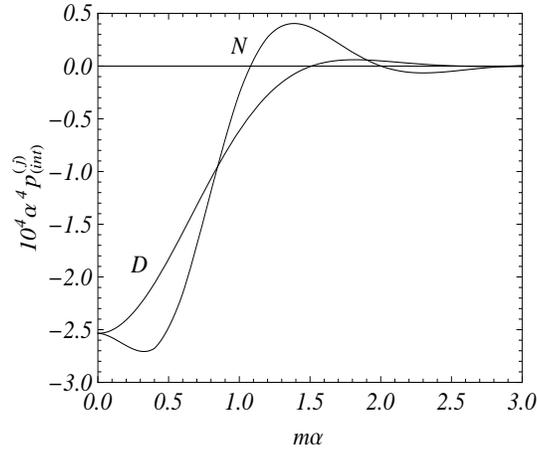,width=7.cm,height=6.cm}
\end{center}
\caption{Interaction force between the plates for $a/\protect\eta =3$ as a
function of the field mass, for a $D=3$ conformally coupled scalar field
with Dirichlet and Neumann BCs. }
\label{fig4}
\end{figure}

In Fig.~\ref{fig4} the dependence of the Casimir force on the parameter $%
m\alpha $ for a given separation corresponding to $a/\eta =3$ is depicted.
Conformally coupled scalar fields with Dirichlet and Neumann BCs,
respectively, are considered.

From the discussion given above it follows that for the proper distances
between the plates larger than the curvature radius of the dS spacetime, $%
\alpha a/\eta \gtrsim \alpha $, the gravitational field essentially changes
the behavior of the Casimir forces compared with the case of the plates in
Minkowski spacetime. In particular, the forces may become repulsive at large
separations between the plates. In particular, for real values $\nu $ and
for Neumann BC on both plates, Casimir forces are repulsive at large
separations, in the range of parameters for which $f_{D}<0$ [see Eqs. (\ref%
{pjlarge}), (\ref{gnuj})]. Recall that, for the geometry of parallel plates
on the background of Minkowski spacetime, the only case with repulsive
Casimir forces at large distances corresponds to Neumann BC on one plate and
non-Neumann BC on the other. A remarkable feature of the influence
of the gravitational field is the oscillatory behavior of the Casimir forces
at large distances, which appears in the case of imaginary $\nu $. In this
case, the values of the plate distance yielding zero Casimir
force correspond to equilibrium positions. Among them, the positions with negative
derivative of the force with respect to the distance are locally stable.

In the discussion above we have considered the expectation values assuming
that the field is prepared in the Bunch--Davies vacuum state. This
corresponds to the effect of vacuum polarization by boundary conditions. In
states containing particles the expectation values of physical observables
will receive additional contributions. For example, the expectation value of
the field squared has the form $\langle \varphi ^{2}\rangle =\langle \varphi
^{2}\rangle _{\text{BD}}+2\sum_{i}\,^{i}n_{\sigma }\varphi _{\sigma
}(x)\varphi _{\sigma }^{\ast }(x)$, where $^{i}n_{\sigma }$ is the number of
particles with the set of quantum numbers $\sigma $ and $\langle \varphi
^{2}\rangle _{\text{BD}}$ is the expectation value in the Bunch--Davies
vacuum state. On the base of this relation, we can consider the effects of
the boundaries at finite temperature assuming that the field is in
thermodynamical equilibrium. In this case, the VEVs are changed by the
thermodynamical expectation values. However, it should be noted that, owing to
the time dependence of the background spacetime, we can talk about 
thermodynamical equilibrium in the adiabatic limit only. In dS spacetime
this corresponds to the conditions $\eta K\gg m\alpha $ and $\eta K\gg 1$,
for the modes with wave number $K$ (see also the discussion in \cite%
{Birr82}). By taking into account that, at temperature $T$, the main
contribution to the thermodynamic expectation values comes from the region
$K\lesssim T$, we obtain the conditions $T_{c}\gg 1/\alpha $ and $T_{c}\gg m$,%
where $T_{c}=\eta T/\alpha $ is the comoving temperature. The dominant
contribution to the boundary induced expectation values comes from the
fluctuations with $K\lesssim 1/a$. Combining this with the estimates given
above, we conclude that the adiabatic approximation for the boundary induced
expectation values corresponds to the limit $a/\eta \ll 1$. As has been
shown above, in this limit the leading terms in the VEVs induced by the
plates coincide with the corresponding quantities for the geometry of
parallel plates in Minkowski spacetime. The same is true for the
thermal corrections.

\section{Conclusion}

\label{sec:Conc}

Amongst the most interesting topics in the investigation of the Casimir
effect is the dependence of the characteristics of the vacuum fluctuations
on the background geometry. In the present paper we have considered the
classical geometry of two parallel plates on the background of dS spacetime
for a scalar field with Robin boundary conditions on the plates. The general
case has been investigated when the constants in the Robin boundary
conditions are different for the two separate plates. In the region between
the plates, the Wightman function has been obtained and displayed under the
form of a mode sum involving series over zeros of the function defined by
Eq.~(\ref{kDvalues}). For the summation of this series we have made use of
expression (\ref{sumfor}). This has allowed us to extract, from the Wightman
function, the part coming from a single plate, and to express the additional
part in terms of integrals, which are exponentially convergent in the
coincidence limit. The single plate contribution was investigated
previously, in Ref.~\cite{Saha09}. The contribution induced by the second
boundary has been presented in two alternative forms, as given by Eqs.~(\ref%
{WF2}) and (\ref{WF3}). By using the expression of the Wightman function, we
have evaluated the VEVs of the field squared and of the energy-momentum
tensor, in the region between the plates. These VEVs are decomposed into a
boundary-free dS, a single plate-induced and an interference contribution,
respectively. The last one, for the cases of the field squared and
energy-momentum tensor, is given by Eqs.~(\ref{phi2Int}) and (\ref{Tll2}), (%
\ref{T0D2}), respectively. The vacuum energy-momentum tensor is
non-diagonal, with the off-diagonal component corresponding to the energy
flux along the direction normal to the plates. In the case of a conformally
coupled massless field, the total single plate contribution to the VEV of
the energy-momentum tensor vanishes and the interference part is obtained
from the corresponding result for the Minkowski bulk, by standard conformal
transformation.

Various limiting cases have been studied. In the limit of small distances
between the plates the interference part in the VEV of the field squared
coincides, to leading order, with the corresponding quantity for a
conformally coupled massless field, and is given by Eq.~(\ref{phi2Conf}).
The leading terms of the interference parts of the VEV for the energy
momentum tensor are given by expressions (\ref{Tllsmall}). For a conformally
coupled scalar field, the leading term of the off-diagonal component
vanishes, and the leading terms of the diagonal components are homogeneous.
In the opposite asymptotic limit of large separations between the plates,
the behaviors of the interference parts crucially depend on the value of the
parameter $\nu $, defined by Eq.~(\ref{nu}). For positive values of this
parameter, the leading terms of the corresponding asymptotic expansions are
given by Eqs.~(\ref{phi2intasr}) and (\ref{TllLarge}), for the field squared
and the energy-momentum tensor, respectively. The interference contributions
for the field squared and the diagonal components of the energy-momentum
tensor decay as $1/(a/\eta )^{D-2\nu }$, whereas the off-diagonal component
decays like $1/(a/\eta )^{D-2\nu +1}$. To leading order, the vacuum stresses
are isotropic and the boundary induced VEV of the energy-momentum tensor
corresponds to a gravitational source of barotropic type, with equation of
state parameter equal to $-2\nu /D$. At large separations between the plates
and for imaginary values of the parameter $\nu $, the asymptotic behavior of
the interference parts for the field squared and for the energy-momentum
tensor are given by Eqs.~(\ref{phi2intasi}) and (\ref{TllLargeIm}),
respectively. The corresponding behavior is damping oscillatory and the VEVs
decay as $(a/\eta )^{-D}\cos [2|\nu |\ln \left( 2a/\eta \right) +\psi ]$,
for the field squared and the diagonal components of the energy-momentum
tensor. For the off-diagonal component the amplitude decays as $(a/\eta
)^{-D-1}$.

The vacuum forces acting on the plates are determined by the $_{D}^{D}$%
-component of the stress. They have been studied in Sect.~\ref{sec:Forces}.
The normal stresses on the plates are presented as sums of single plate and
interaction contributions. The contributions to the vacuum force coming from
the single plate terms are the same from the left and from the right sides
of the plate and thus give no contribution to the effective force. The
interaction forces per unit surface are determined by formula (\ref{pintj}).
This expression is further simplified in the special cases of Dirichlet and
Neumann BCs, yielding Eqs.~(\ref{pjD}) and (\ref{pjN}), respectively. For
small distances between the plates, to leading order the vacuum forces are
given by Eq.~(\ref{pintjsmall}). If, in addition, $|\beta _{j}|/a\gg 1$, the
vacuum forces are attractive at small distances, except for the case of
Dirichlet BC on one plate and non-Dirichlet on the other, in which case the
force turns out to be repulsive. At large distances between the plates and
for positive values of $\nu $, the force acting on the plate decays
monotonically as $1/(2a/\eta )^{D-2\nu +2}$, for non-Neumann BCs, and as $%
1/(2a/\eta )^{D-2\nu }$, in the case of Neumann BCs [see Eqs.~(\ref{pjlarge}%
)]. For imaginary values of $\nu $ the behavior of the vacuum forces is
damping oscillatory, in the leading order described by Eqs.~(\ref{pjlargeim}%
). Having in mind that spectral properties of spin 2, 1, 0 operators for dS
spacetime are known, the current study can be now extended to the
calculation of the Casimir force due to quantum gravity (for the one-loop
effective action of arbitrary quantum gravity in dS spacetime see Ref.~\cite%
{Cogn05}). This will be considered elsewhere.

From the analysis carried out above, it follows that the curvature of the
background spacetime decisively influences the behavior of boundary induced
VEVs at distances larger than the curvature scale. As we have seen, when the
background is dS spacetime the decay of the VEVs at large separations
between the plates is power-law (monotonical or oscillating), independently
of the field mass. This is quite remarkable and clearly in contrast with the
corresponding features of the same problem in a Minkowski bulk. To wit, the
interaction forces between two parallel plates in Minkowski spacetime at
large distances decay as $1/a^{D+1}$ for massless fields and these forces
are exponentially suppressed for massive fields by a factor of $e^{-2ma}$.
For the geometry of two parallel plates, in AdS spacetime the decay of the
vacuum forces at large separations is also exponential (see Ref.~\cite%
{Saha05}), with the suppression factor being determined by the AdS curvature
scale. In a way very much similar to the procedure described in Ref.~\cite%
{Eliz09} (see also Ref.~\cite{Teo09} for finite temperature effects), we are
able to treat here the more general case of dS spacetime with compact
internal subspaces and piston-like geometries. Note that this calculation
can be extended to a self-interacting scalar field theory too, in which case
mass becomes an effective mass, proportional to the background scalar. In
this way, our results and procedures here can be used to the study of
curvature-induced phase transitions of the in-in effective potential in the
same way as it was proposed for the out-in effective potential in Ref.~\cite%
{Buch85}.

\section*{Acknowledgments}

The authors are grateful to Sergei Odintsov for helpful discussions and
comments. Part of EE's research was performed while on leave at Department
of Physics and Astronomy, Dartmouth College, 6127 Wilder Laboratory,
Hanover, NH 03755, USA. EE was supported by MICIIN (Spain), project
FIS2006-02842, and by AGAUR (Generalitat de Ca\-ta\-lu\-nya), contract
2005SGR-00790 and grant DGR2009BE-1-00132. AAS was supported by the ESF
Programme ``New Trends and Applications of the Casimir Effect" and in part
by the Armenian Ministry of Education and Science Grant No. 119.

\appendix

\section{Minkowski spacetime limit}

\label{sec:AppMink}

In this appendix we will explicitly show the limiting transition of the
situations considered above to the geometry of two parallel Robin plates in
Minkowski spacetime, for the vacuum interaction forces. In this limit $%
\alpha \rightarrow \infty $ and the modulus of the order of the modified
Bessel functions is large, $\nu \approx im\alpha $. In addition, we have $%
\eta \rightarrow \alpha $. Changing the integration variables to $x=u\eta $,
$y=v\eta $, we see that the arguments of the modified Bessel functions are
large, too. Hence, we make use of the uniform asymptotic expansions for
these functions for imaginary values of the order with large modulus. The
leading terms in these expansions have the form (see, for example, \cite%
{Milt09})%
\begin{eqnarray}
&&K_{i\mu }(\mu z)\sim \sqrt{2\pi /\mu }e^{-\mu \pi /2}\cos [\mu f(z/\mu
)-\pi /4],  \notag \\
&&I_{i\mu }(\mu z)+I_{-i\mu }(\mu z)\sim -\frac{2e^{\mu \pi /2}}{\sqrt{2\pi
\mu }}\sin [\mu f(z/\mu )-\pi /4],  \label{IKas1}
\end{eqnarray}%
for $z<1$ and
\begin{eqnarray}
&&K_{i\mu }(\mu z)\sim \sqrt{\frac{\pi }{2\mu }}\frac{e^{-\mu \pi /2}}{%
(z^{2}-1)^{1/4}}e^{-\mu g(z)},  \notag \\
&&I_{i\mu }(\mu z)+I_{-i\mu }(\mu z)\sim \frac{2}{\sqrt{2\pi \mu }}\frac{%
e^{\mu \pi /2}}{(z^{2}-1)^{1/4}}e^{\mu g(z)},  \label{IKas2}
\end{eqnarray}%
for $z>1$. The functions in these formulas are defined as
\begin{eqnarray}
f(z) &=&\ln \left( \frac{1+\sqrt{1-z^{2}}}{z}\right) -\sqrt{1-z^{2}},  \notag
\\
g(z) &=&-{\mathrm{arcsec\,}}z+\sqrt{z^{2}-1}.  \label{fgz}
\end{eqnarray}%
From Eqs.~(\ref{IKas1}) and (\ref{IKas2}) it follows that%
\begin{eqnarray}
\tilde{I}_{\nu }(y)\tilde{K}_{\nu }(y) &\sim &\frac{y^{D}}{m\alpha }\cos
[2m\alpha f(y/m\alpha )],\;y<m\alpha ,  \notag \\
\tilde{I}_{\nu }(y)\tilde{K}_{\nu }(y) &\sim &\frac{1}{m\alpha }\frac{y^{D}}{%
\sqrt{(y/m\alpha )^{2}-1}},\;y>m\alpha .  \label{IKtildeas}
\end{eqnarray}%
The main contribution to the force (\ref{pintj}) comes from the region $%
(m\alpha ,\infty )$ of the integration over $y$; to leading order, we find
\begin{equation}
p_{\text{(int)}}^{(j)}\approx -\frac{2(4\pi )^{-D/2}}{\Gamma (D/2)}%
\int_{m}^{\infty }du\,\,\frac{u^{2}(u^{2}-m^{2})^{D/2-1}}{\frac{(\beta
_{1}u-1)(\beta _{2}u-1)}{(\beta _{1}u+1)(\beta _{2}u+1)}e^{2au}-1}.
\label{pjMink}
\end{equation}%
This result coincides with the corresponding formula for parallel plates in
the Minkowski bulk. In a similar way, the limiting transition for the other
quantities can also be checked.

\end{document}